\documentclass{basi}
\usepackage{graphicx}
\def\lsim{\lower.5ex\hbox{$\; \buildrel < \over \sim \;$}}
\def\gsim{\lower.5ex\hbox{$\; \buildrel > \over \sim \;$}}
\def \simeq{\lower.3ex\hbox{$\; \buildrel \sim \over - \;$}}
\def\ch{\lower-0.55ex\hbox{--}\kern-0.55em{\lower0.15ex\hbox{$h$}}}
\def\lh{\lower-0.55ex\hbox{--}\kern-0.55em{\lower0.15ex\hbox{$\lambda$}}}

.tfm

\begin{document}

\title [Nature of GRBs observed by RT-2 onboard CORONAS-PHOTON Satellite]
{Nature of GRBs observed by RT-2 onboard CORONAS-PHOTON Satellite}

\author [D. Debnath, A. Nandi, S. K. Chakrabarti, T. B. Katoch, and A. R. Rao]
{D. Debnath$^{1}$, A. Nandi$^{2}$, S. K. Chakrabarti$^{1,3}$, T. B. Katoch$^{1,4}$, A. R. Rao$^{4}$\\
$^1$ Indian Centre for Space Physics, Chalantika 43, Garia Station Rd., Kolkata, 700084, India\\
$^2$ Space Astronomy Group, ISAC, HAL Airport Road, Bangalore, 560017, India\\
$^3$ S. N. Bose National Center for Basic Sciences, JD-Block, Salt Lake, Kolkata, 700098, India\\
$^4$ Tata Institute of Fundamental Research, Homi Bhabha Road, Mumbai, 400005, India\\
e-mail: dipak@csp.res.in,  anuj@isac.gov.in, chakraba@bose.res.in, tilak@tifr.res.in, arrao@tifr.res.in }

\maketitle

\begin{abstract}
\noindent {The RT-2 Experiment, a low energy gamma-ray telescope, onboard CORONAS-PHOTON satellite 
is designed to study the temporal, spectral and spatial properties of the hard X-ray solar flares 
mainly in the energy range of $15 - 100$ keV, which is extendable upto $1000$ keV. During the
operational period of $\sim 9$ months, it has been able to detect a few solar flares and at least four 
Gamma-Ray Bursts (GRBs). In this paper, we discuss the properties of the GRBs as observed 
by RT-2 in the energy band of 15 to $\sim$1000 keV. 
%We will also augmented these results with the Swift results. 
We will present the results of spectral and timing properties of the GRBs (specially for the 
GRB 090618) using RT-2 detectors. Temporal analysis suggests that all four GRBs belong to the 
category of long duration bursts.

%In GRB 090618, multiple peaks were observed, and we also found that the peak energies ($E_p$) was 
%varied in monotonically decreasing manner from one peak to another. 
}

\end{abstract}

\vskip 0.2cm
\begin{center}
\end{center}

%\end{document}

\section{Introduction}

RT-2 (named after R\"{o}ntgen) Experiment (Debnath et. al. 2011, Nandi et. al. 2011,
Kotoch et. al. 2011, Sarkar et. al. 2011, Sreekumar et. al. 2011, Rao et al. 2010, 2011a)
is a part of Russian Solar mission satellite named Coronas-Photon. It was launched 
into $\sim$550~Km polar LEO (Low Earth Orbit) on 30$^{th}$ January, 
2009 from Plesetsk Cosmodrome, Russia. The mission was dedicated to study mainly the solar 
flares in a wide energy band of electromagnetic spectrum ranging from UV to high-energy 
$\gamma$-rays ($\sim$2000 MeV). But the mission has lost its functionality 
after $\sim 9$ months of its successful launch due to the  failure of the main power-bus of the 
satellite.

RT-2 is an Indian Space Research Organization (ISRO) sponsored X-ray and low-energy $\gamma$-ray
(up to 1 MeV) solar experiment in collaboration with MEPHI, Russia. It consists of an ensemble of 
three low-energy hard X-ray/gamma-ray payloads (RT-2/S, RT-2/G \& RT-2/CZT). RT-2/S \& RT-2/G are 
mainly sensitive in the energy range of $15 - 100$~keV. They also have extended detection 
capabilities, up to $\sim 600$~keV for RT-2/S and up to $\sim 1000$~keV for RT-2/G 
(Debnath et al. 2011). But, RT-2/CZT is sensitive to the spectral energy range of $20 - 100$~keV 
with imaging capability.

The main scientific objectives of the RT-2 experiment were to explore temporal, spectral and 
spatial behaviors of the Sun and solar events, such as solar flares etc. Due to the large FOV of the 
RT-2 instruments and having the capability to act as an omni-directional detector in high energy 
range ($> 50$ keV), detectors have been able to detect a few extra-solar events such as 
Gamma-Ray Bursts (GRBs). During the mission life time, RT-2 has co-discovered four GRBs - GRB 090618 
(Rao et al. 2009, 2011a), GRB 090820, GRB 090926A, and GRB 090929A 
(Chakrabarti et al. 2009a, 2009b, 2009c).

In this {\it paper}, we discuss the timing and spectral properties of the first observed GRB 090618 and 
preliminary results of the other GRBs. The paper has been organised in the following way: in \S 2, 
a brief discussion on RT-2 observations and data analysis procedures is given, in \S 3 
observational results of the GRBs are discussed and in \S 4 discussion and concluding remarks.

\section {Observation and Data Analysis}

In this {\it paper}, we present the results of the four GRBs as observed by RT-2. During its 
operational period, RT-2 has successfully co-discovered GRB 090618, GRB 090820, GRB 090926A \& 
GRB 090929A. 
All these four GRBs were detected by RT-2/S and RT-2/G. However, RT-2/CZT has not been 
able to image any of these GRBs, because of the long duration commissioning phase of RT-2/CZT 
operation and the entire payload is shielded with Ta material, which restricts the detection of 
GRBs as these were observed at high inclination angle. In this Section, we will discuss about the 
observations and the data analysis procedure in briefly. 
%We also discuss about the timing and the spectral properties of the GRBs as observed by RT-2. 
%We also augment our results with the results obtained from publicly available Swift data, s
%pecially for the first GRB.

RT-2/S and RT-2/G are generally operated in the solar quiet mode (SQM) when the satellite orbits 
in the``GOOD" regions (i.e., outside the high background regions of Polar Caps and SAA), when
count rates in eight channels (for each detector) are stored every second. The spectral data 
are stored every 100 s. The low energy spectra are stored separately for NaI(Tl) ($15 - 100$ keV)
and NaI(Cs) ($25 - 215$ keV) detectors based on the pulse shape along with the high-energy 
spectrum in the energy range of $215 - 600/1000$~keV (see, Debnath et al. 2011, for details). 
The onboard software automatically stores the data in finer time resolution ($0.1$ s count rates 
and $10$ s spectra) during the solar flare mode (SFM), when the count rates exceed a 
pre-determined threshold limit. RT-2/CZT operates only in the SQM when $1$ s count rates,
$100$ s spectra, and images are stored. It has no solar flare operation mode. 
The onboard performance of the RT-2 instruments are discussed in Nandi et al. (2009) and 
Rao et al. (2011b).

%The test and evaluation results of these payloads are described in details in 
%Debnath et al. (2011), Kotoch et al. (2011), Nandi et al. (2011), Sarkar et al. (2011), and 
%Sreekumar et al. (2011).

RT-2 light curve and spectral data were extracted from the `raw' data using ground-based software. 
%using flight-model (FM) test softwares, written in LabVIEW. 
The data is processed further with the NASA's software package heasoft6.8 using FTOOLS and XSPEC.
%Spectral data were fitted by using the XSPEC package.

\section {Results}

\subsection{GRB 090618}

The GRB 090618 was discovered with the Swift BAT on 2009 June 18 at 08:28:29 UT 
(Schady et al. 2009) at a redshift of $z = 0.54$ (Cenko et al. 2009a). The temporal profile of 
Swift/BAT showed that the GRB was very intense with multi-peak emission in gamma-rays. The GRB was 
also detected by various observatories in X-ray and gamma-ray energies such as AGILE 
(Longo et al. 2009), Fermi GBM (McBreen et al. 2009), Suzaku WAM (Kono et al. 2009), 
Konus-Wind on board the Wind satellite and Konus-RF on board the Coronas-Photon satellite 
(Golenetskii et al. 2009a), the RT-2 Experiment on board the Coronas-Photon satellite 
(Rao et al. 2009), etc. The optical afterglow of GRB 090618 was also immediately detected 
with the ROTSE-IIIb (Rujopakarn et al. 2009), Palomar 60 inch telescope (Cenko et al. 2009b). 
The Swift X-ray Telescope (XRT) was able to detect the X-ray afterglow (very bright in X-rays), 
just 121 sec after the BAT trigger. Within a few secs, the X-ray flux decayed rapidly
with a slope of $\sim 6$ before breaking at $T_0 + 310$ s ($T_0 = 08:28:29$ UT) to a almost 
flatter slope of $0.71 \pm 0.02$ (Beardmore et al. 2009). 

GRB 090618 was very intense in hard X-rays and gamma-rays during the prompt emission phase 
that makes enable to study the time-resolved spectral characteristics of the different peak
profiles. 
%Significant spectral evolution was observed during the prompt emission of the burst. 
%Time-averaged spectrum from $T_0 - 4.4$ to $T_0 + 213.6$ s was well fitted by a power law 
%with an exponential cutoff with the photon index of $1.42$ and $E_p$ of $134$ keV 
%(Sakamoto et al. 2009). 
The GRB (band) model fitted spectra in the energy range of $20$ keV - $2$ MeV that obtained from 
%the Konus-Wind (from $T_0$ to $T_0 + 142$ s; 
%$T_0 = 08:28:24.974$ UT) on board the Wind satellite and 
the Konus-RF (from $T_0$ to $T_0 + 142$ s; $T_0 = 08:28:27.060$ UT) detector on board the 
Coronas-Photon satellite, provided the important parameters of the 
low-energy photon index ($\alpha$), high-energy photon index ($\beta$), 
and peak energy ($E_p$) as $-1.28$, $-3.06$, and $220$ keV (Golenetskii et al. 2009a). 
The BAT light curve of the GRB in the energy band of 15-150 keV was found to be of a multi-peak 
structure with a duration of about $130$ s.
%The time-averaged BAT spectrum from $T_0 - 5$ to $T_0 + 109$ s can be described by simple 
%power-law model with index $\sim 1.7$ (Baumgartner et al. 2009). The fluence in the 
%$15-100$ keV band is $1.06 \pm 0.01 \times 10^{-4}~erg~cm^{-2}$. 
The multi-peaked profile was also observed in the $50$ keV - $5$ MeV range light curve of 
the Suzaku Wide-band All-sky Monitor (Kono et al. 2009).

\begin{figure}[h]
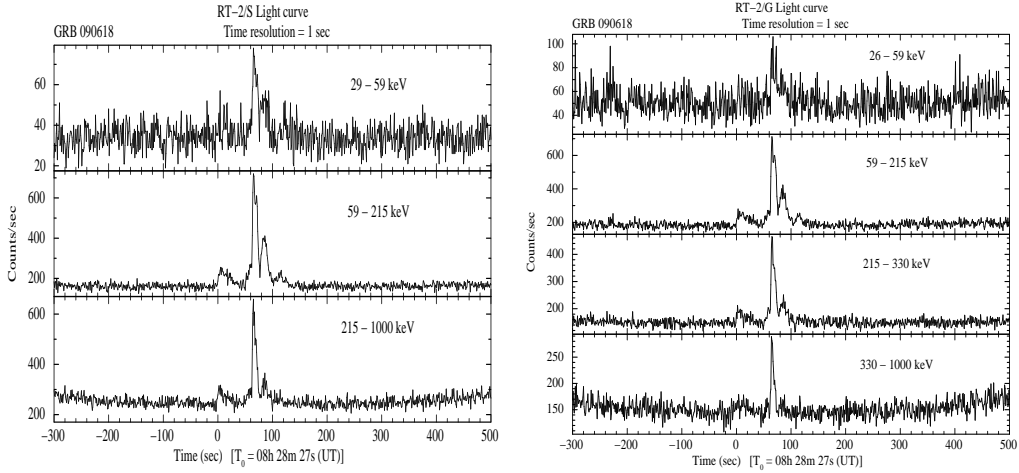

\centering
\includegraphics[height=2.6in,width=2.5in,angle =270]{grb090618-s-lc.ps}\hspace{0.1 cm}
\includegraphics[height=2.6in,width=2.5in,angle=270]{grb090618-g-lc.ps}
\caption{(a-b) Light curves of GRB~$090618$ in different energy band (marked) of (a) RT-2/S and 
(b) RT-2/G detectors. Four emission peaks along with the {\it ignition (precursor trigger) pulse} 
at T$_0$ = 08h 28m 27s UT are detected clearly in the energy band of 59 - 215 keV.}
\label{kn : fig1}
\end{figure}

RT-2 instruments detected the GRB 090618 with a large off-axis angle of $77^\circ$, which 
triggered at $T_0 = 08:28:27 UT$ and lasted for almost 150 sec. During the event, the RT-2 
payload was completely in the ``SHADOW" mode (away from the Sun) with ``GOOD'' time (away from 
SAA and polar regions), which started at $08:16:10.207$ UT and ended at $08:37:35.465$ UT. 
Data from RT-2/S and RT-2/G are used for the present analysis. 

\subsubsection{Timing analysis results for GRB 090618}

GRB 090618 was detected in the different energy bands of RT-2/S and RT-2/G detectors.
The 1 sec binned light curves of the GRB, as seen by RT-2/S and RT-2/G are shown in Figure 1.
The multi-structured burst profile is clearly seen in the light curve of 59 - 215 keV. The
emission peaks are registered at $T_0$+65 s, $T_0$+85 s and at $T_0$+115 s. 
To find the widths of the peaks, we tried to fit RT-2/G light curve data of 59-215 keV energy 
band with fast-rise-exponential-decay (FRED) profile developed by Kocevski et al. (2003). 

\begin{figure}[h]
\centering
\includegraphics[height=2.4in,width=2.8in,angle=000]{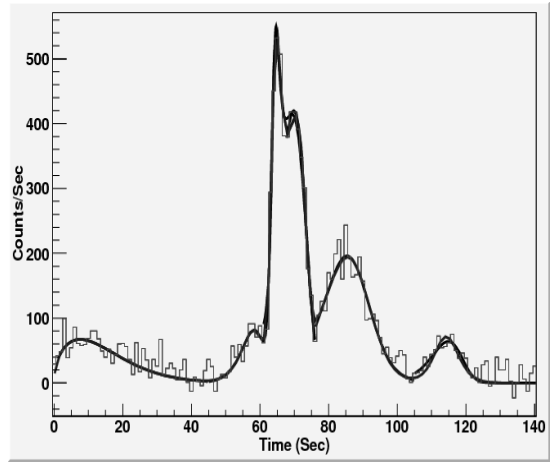}
\caption{Multiple FRED profiles fitted for the background subtracted light curve of RT-2/G in the
energy band of 59 - 215 keV ($T_0$ = 08h 28m 27s UT) for GRB 090618.}
\label{kn : fig2}
\end{figure}

According to the FRED profile GRB photon flux can be defined as:
$$
F(t) = F_m(\frac {t}{t_m})^r[\frac {d}{d+r} + \frac {r}{d+r}
        (\frac {t}{t_m})^{(r+1)}]^{-(r+d)/(r+1)},
\eqno (1)
$$
where $F_m$ is the maximum flux at time $t_m$, $r$ and $d$ are the decaying indices for rising
and decaying phases of any pulse (peak) respectively. We first fit individual peaks and then
total light curve with the pulse profile of Eqn. (1). This model fitted light curve for RT-2/G
detector in the energy band of 59 - 215 keV is shown on Fig. 2. We also fitted this model
in the other energy bands of RT-2/G light curves. Finally we calculated widths of each peaks
from our fittings. All the fitted parameters and pulse widths are noted in Table 1. We also
compared our fitted results with Swift BAT data (Rao et al. 2011a).

\begin{table}[h]
\small
\centering
\caption{\label{table1} FRED profile fitted results for RT-2/G light curves of GRB 090618}
\vskip 0.2cm
\begin{tabular}{|l|c|c|c|c|c|c|}
\hline
 Pulse & Energy range (keV) & $F_m$ (s$^{-1}$) & $t_m$ (s) & $r$ (s)  & $d$ (s)  & Width (s) \\
\hline
{1} &        &  532.0 & 65.0 & 81.1 & 13.8 & 6.30 \\
{2} & 59-215 &  422.0 & 70.0 & 34.5 & 34 & 7.15  \\
{3} &        &  210.0 & 85.0 & 20.0 & 20.0 & 14.60 \\
{4} &        &  72.0 & 114.0 & 42.0 & 40.0 & 9.49\\
\hline
{1} &          & 315.0 & 65 & 82.5 & 16.4 & 5.59 \\
{2} & 215-330 & 200.0 & 70.0 & 41.6 & 34.2 & 6.17 \\
{3} &         & 76 & 85.0 & 25.0 & 22.0 & 12.17 \\
\hline
{1} & 330-1000 & 144.0 & 65.0 & 78.3 & 22.5 & 4.77  \\
{2} &        &  64.0 & 70.0 & 88.0 & 84.0 & 2.93  \\
\hline
\end{tabular}
\end{table}

\subsubsection{Spectral analysis results for GRB 090618}

Spectral analysis for the GRB was done by using the RT-2/S and Swift BAT data. We used
$15-200$ keV BAT, $100-210$ keV $RT$-2/S G1, and $250-600$ keV $RT$-2/S G2 data for
our combined spectral fitting analysis.
While fitting the combined spectrum of BAT and RT-2, the relative normalisation factor was 
kept fixed at 1.30. Detailed spectral analysis is given in Rao et al. (2011a).
The combined spectrum is well fitted with the standard BAND model.
Spectral fitted results showed that the time averaged peak energy is at $\sim 164$ keV and 
integrated fluence in $20 - 600$ keV energy band is equal to $2.8 \times 10^{-4}$ ergs/cm$^2$.

\begin{figure}[h]
\centering
\includegraphics[height=2.8in,width=2.4in,angle=270]{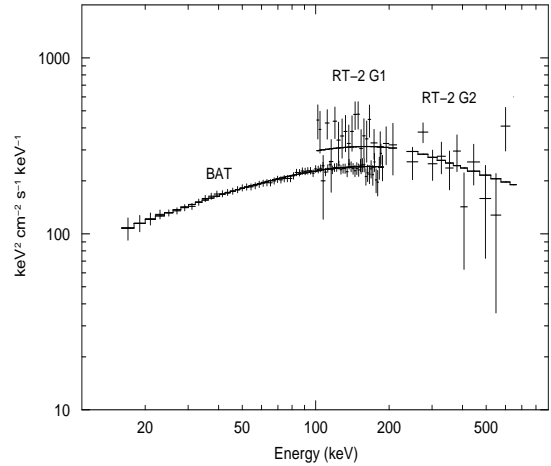}
\caption{The unfolded $Swift$ BAT and $RT$-2/S combined energy spectrum of GRB 090618
(Rao et al. 2011a).}
\label{kn : fig3}
\end{figure}

%For RT-2 spectral data fitting we first generated response matrix using FTOOLs {\it genrsp} task. 
The GRB spectrum can be well fitted with the double power-law model introduced by Band
et al. (1993). In this model, two power-laws join at break frequency $(\alpha-\beta)$E$_0$,
where $\alpha$ and $\beta$ are the first and second power-law indices respectively, and $E_0$ 
is the break energy. Now one can calculate the peak energy $E_p$ by the relation 
$E_p = E_0(2+\alpha)$.
Figure 3 shows the unfolded BAND model fitted $Swift$ BAT and $RT$-2/S combined
energy spectrum of the GRB 090618. 

%{\bf For this combined spectral fit of BAT and RT-2,
%the relative normalization factor was kept fixed at value $1.30$.}

%From the spectral fittings, we measured time averaged peak energy ($E_p$) for the entire 
%burst is $\sim 164$ keV.

\begin{figure}[h]
\centering
\includegraphics[height=2.6in,width=2.5in,angle=270]{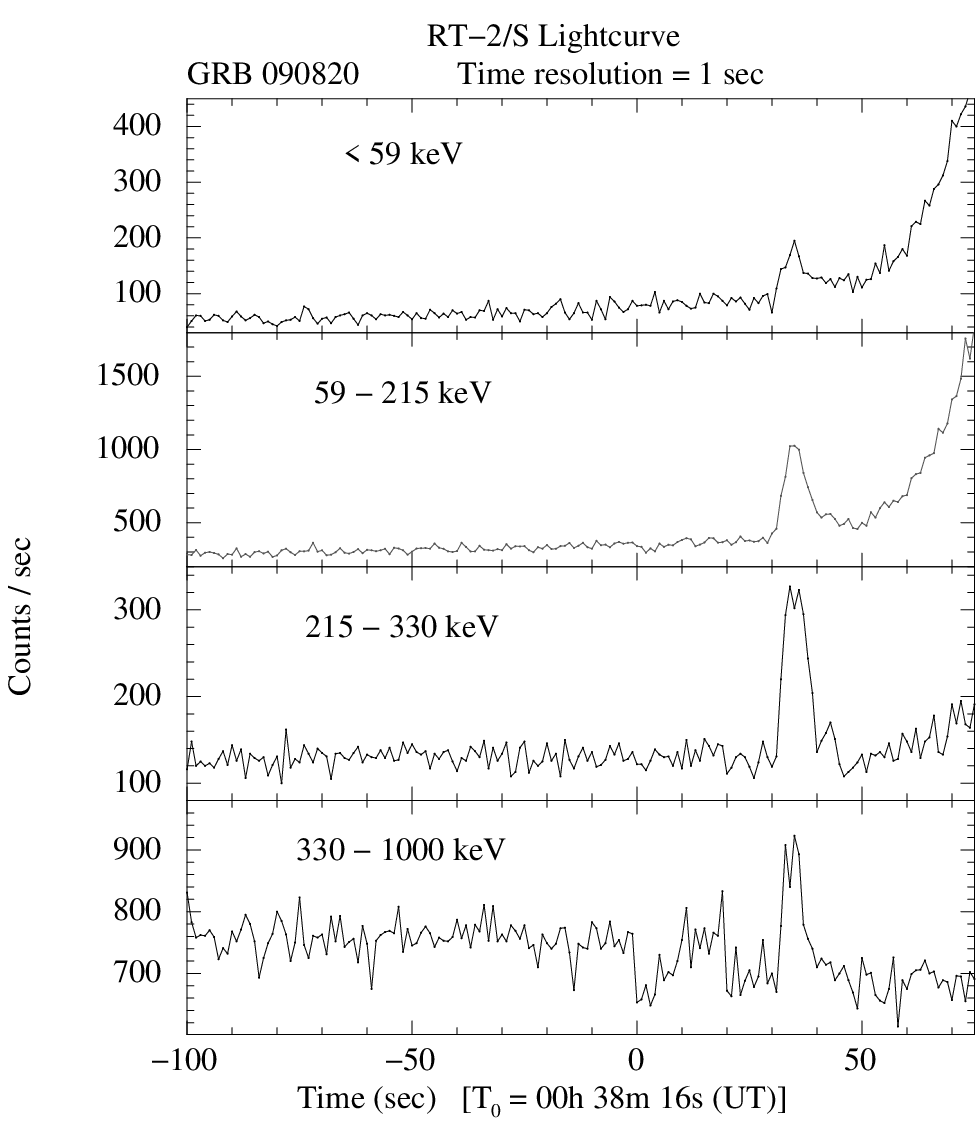}\hspace{0.1 cm}
\includegraphics[height=2.6in,width=2.5in,angle=270]{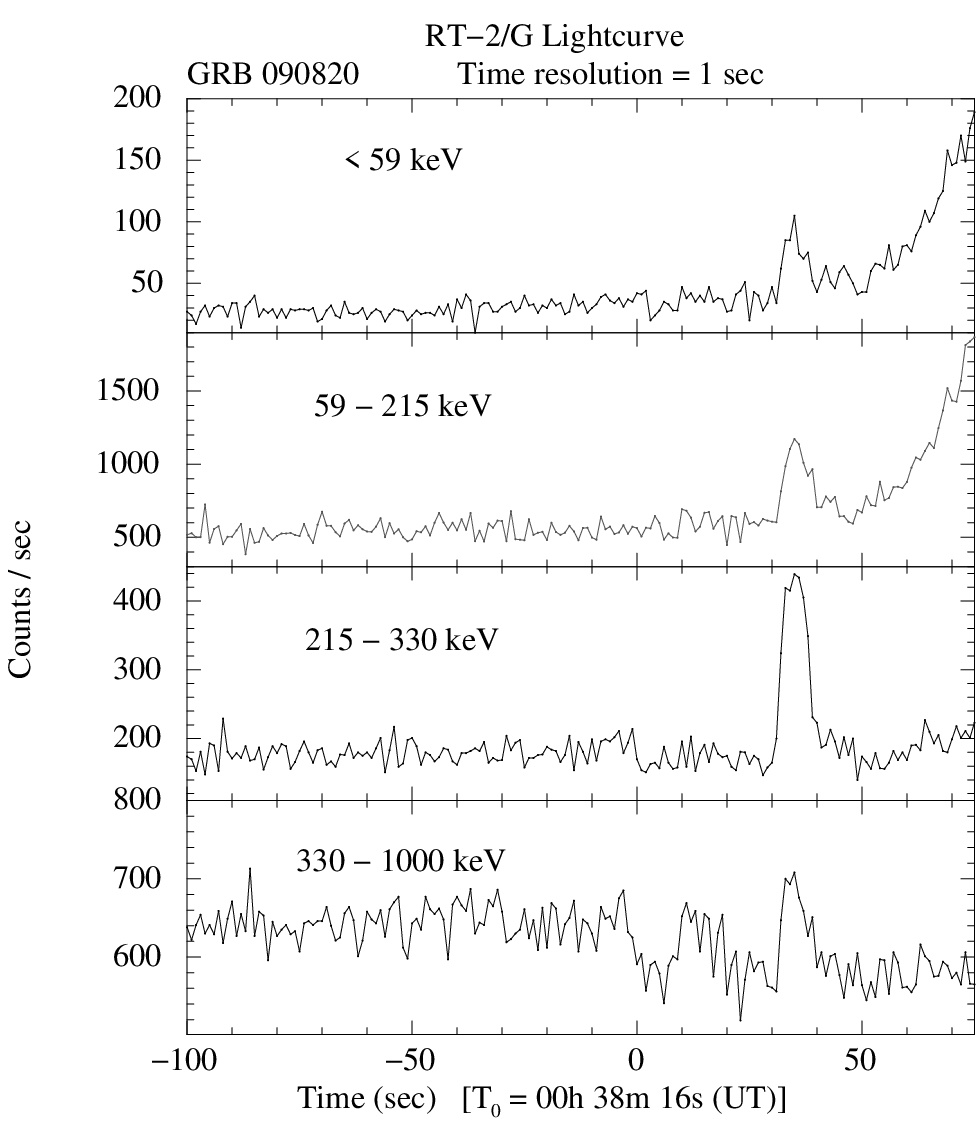}
\caption{(a,b) Light curves in different energy band (marked) of (a) RT-2/S and (b) RT-2/G
detectors. Emission peaks along with the weak precursor trigger pulse at T$_0$ = 00h 38m 16s UT
are shown (Chakrabarti et al., 2009a).}
\label{kn : fig4}
\end{figure}

\subsection{GRB 090820}

The Fermi Gamma-Ray Burst Monitor (GBM) first detected the GRB 090820 by its trigger
number 272421498 (Connaughton V. 2009) on 20th August, 2009 at 00:38:16.19 UT.
The source location was at RA(J2000) = 87.7 degrees (5h 51m) and Dec(J2000) = 27.0
degrees ($+27^\circ 0^{'}$). The light curve of this bright GRB shows GBM triggered
on a weak precursor, followed by a bright pulse beginning 30 sec later and lasting a
further 30 seconds. The integrated event fluence is around $6.6 \times 10^{-5}$ ergs/cm$^2$ in 
the energy band of 8-1000 keV.

Both the RT-2 phoswich detectors (RT-2/S \& RT-2/G) have detected this bright astrophysical 
event (Chakrabarti et al., 2009a). The satellite was in the GOOD/LIGHT mode with Earth 
latitude = $-64.09$ and longitude = $169.11$ with 600 sec duration of GOOD time observation, 
starting at 00h 29m 16sec (UT) and ending at 00h 39m 16sec (UT). After 30 sec of the event, 
the satellite completely entered into the BAD mode of high charge particle region.

The 1 sec binned light curves of RT-2 detectors are shown in Figure 4. The GRB light curve shows 
a simple profile of a strong emission peak at T$_0$+34 sec with a weak precursor 
(T$_0$ = 00:38:16 UT). The time duration of the brightest emission is around $19$ sec with 
$\sim 1100$ counts/sec in the energy band of 59 - 215 keV.

%Both RT-2/S and RT-2/G detectors have registered this burst profile in the
%energy band of $15 - 1000$ keV with the strongest emission in the energy range of
%around $100$ keV to $330$ keV. It is also noted that the burst width decreases
%with the increase in the energy band.

\subsection{GRB 090926A}

The very bright GRB 090926A was first detected by the Fermi-GBM trigger 275631628
(Bissaldi 2009) on 26th September, 2009 at 04:20:26.99 UT. The source location
was at RA(J2000) = $354.5$ degrees (23h 38m), DEC(J2000) = $-64.2$ degrees
($-64^\circ 12^{'}$), with an uncertainty of $1^\circ$. The GBM light curve
consists of single pulse with a duration (T90) of 20 sec. The integrated fluence in
the energy range of 10 keV to 10 GeV is around $2.47 \times 10^{-4}$ ergs/cm$^2$. 

\begin{figure}[h]
\centering
\includegraphics[height=2.6in,width=2.5in,angle=270]{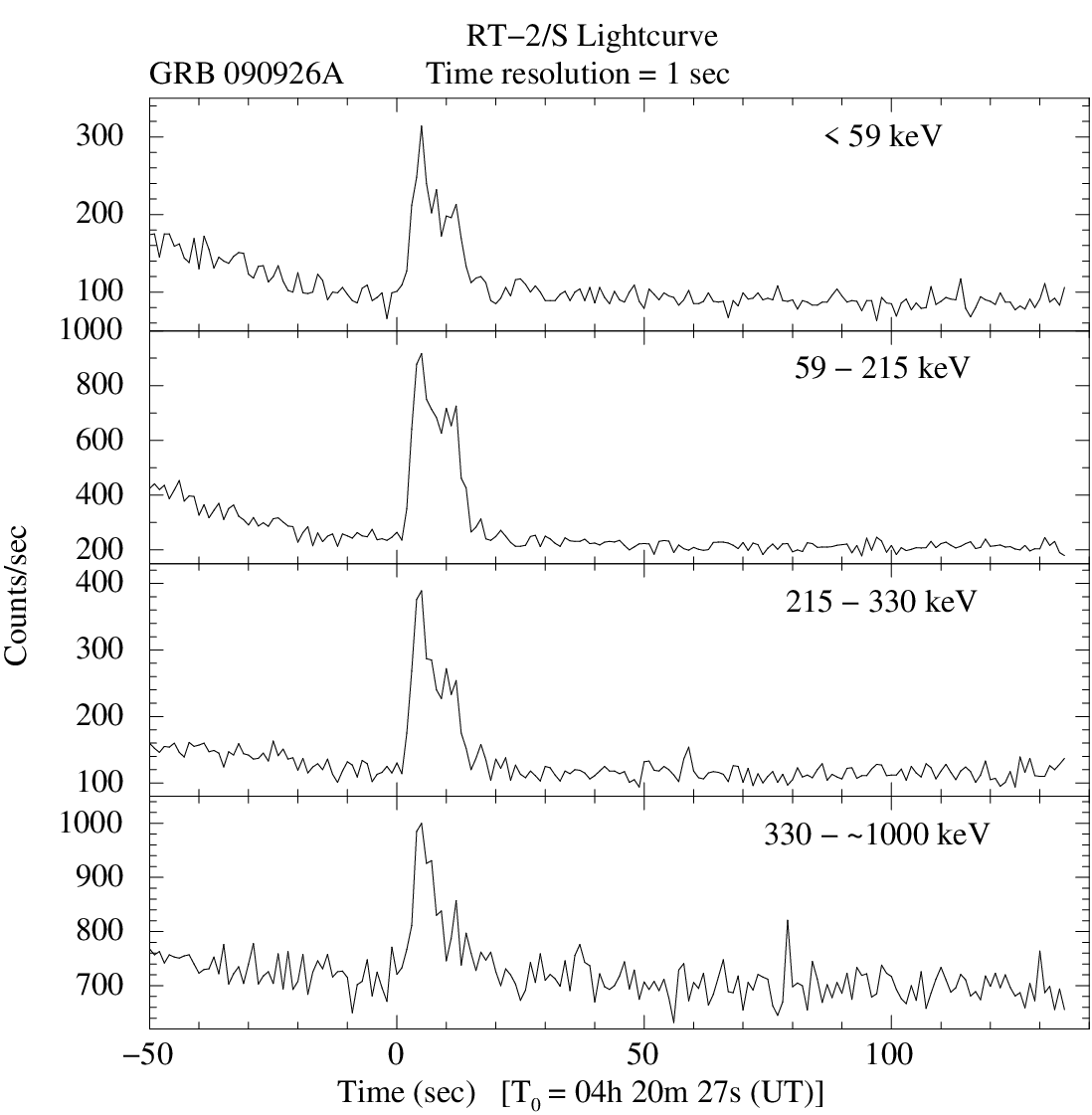}\hspace{0.1 cm}
\includegraphics[height=2.6in,width=2.5in,angle=270]{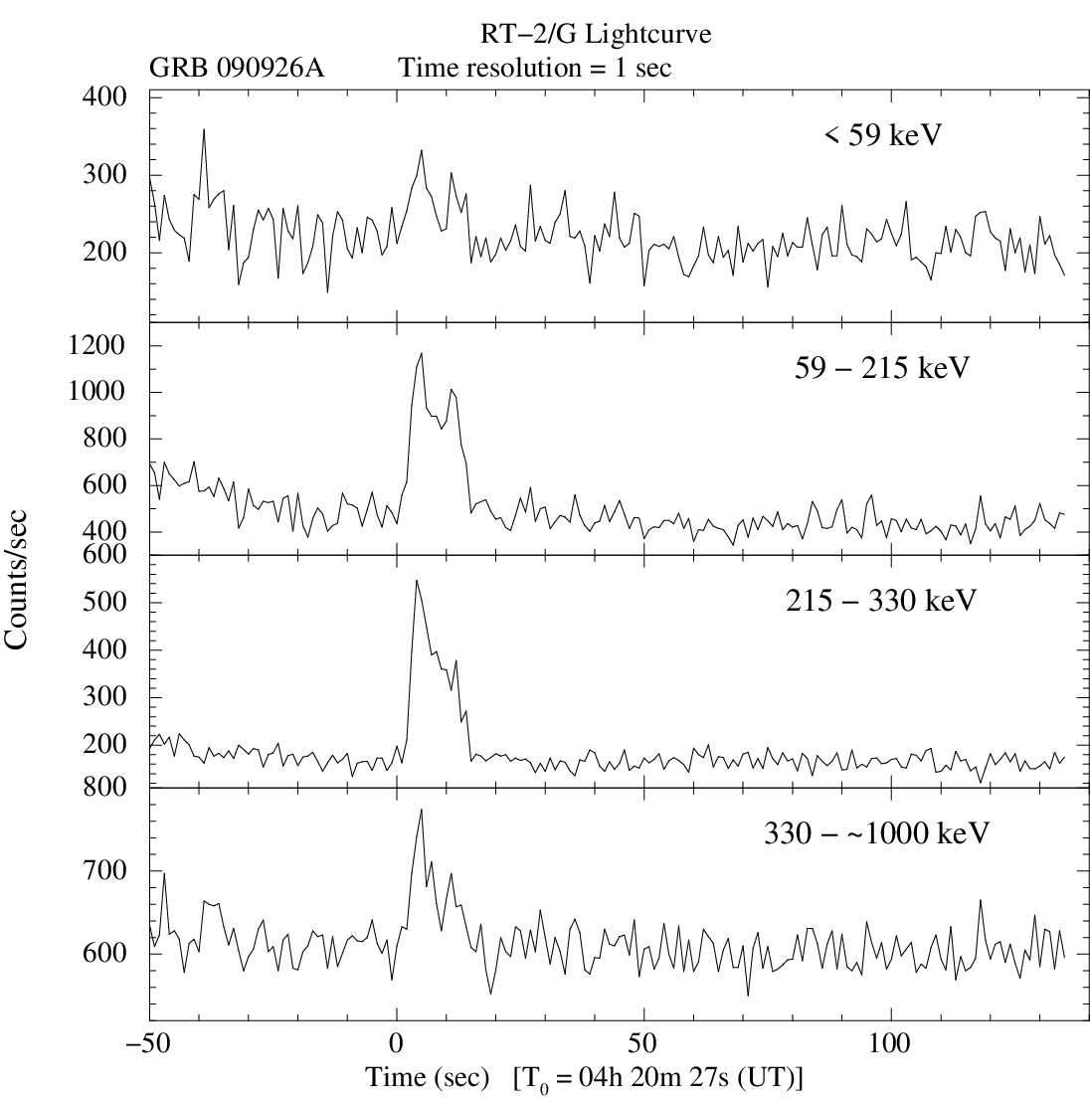}
\caption{(a,b) Light curves in different energy band (marked) of (a) RT-2/S and (b) RT-2/G
detectors. Two emission peaks along with the weak precursor trigger pulse at 
T$_0$ = 04h 20m 27s UT are detected in a wide energy band of 15 - 1000 keV 
(Chakrabarti et al., 2009b).}
\label{kn : fig5}
\end{figure}

RT-2 instruments (RT-2/S \& RT-2/G) also detected this bright GRB at T$_0$ = 04h 20m 27s
(UT) (Chakrabarti et al., 2009b). The satellite was in the LIGHT mode (pointing towards the Sun)
for a short duration at a high latitude in its orbit. During this time, the GOOD time (away from
the polar and SAA regions) observation was for 348 sec starting at 04h 16m 55sec (UT) and ending
at 04h 22m 43sec (UT). This burst was also detected by KONUS-RF, another instrument onboard 
CORONAS-PHOTON satellite (Golenetskii et al., 2009b).

The burst light curve consists of multiple peaks of total duration of $\sim 17$ sec,
followed by a weak tail ending at T$_0$+30 sec. The strongest peak count rate is $\sim 1200$
counts/sec in the energy band of 59- 215 keV. The 1 sec binned light curves of both the
detectors at different energy band are shown in Figure 5.

\subsection{GRB 090929A}

The Fermi Gamma-Ray Burst Monitor first reported the GRB 090929A by its trigger $275891585$
(Rau, A., 2009). The event occurred at 04:33:03.97 UT on 29th September, 2009
in the sky location of RA(J2000) = 51.7 degrees (03h 27m), DEC(J2000) = -7.3 degrees
($-7^\circ 18^{'}$), with an uncertainty of 1.3 degrees. The integrated fluence in the energy
range of 8-1000 keV is around $1.06 \times 10^{-5}$ ergs/cm$^2$.

\begin{figure}[h]
\centering
\includegraphics[height=2.6in,width=2.5in,angle=270]{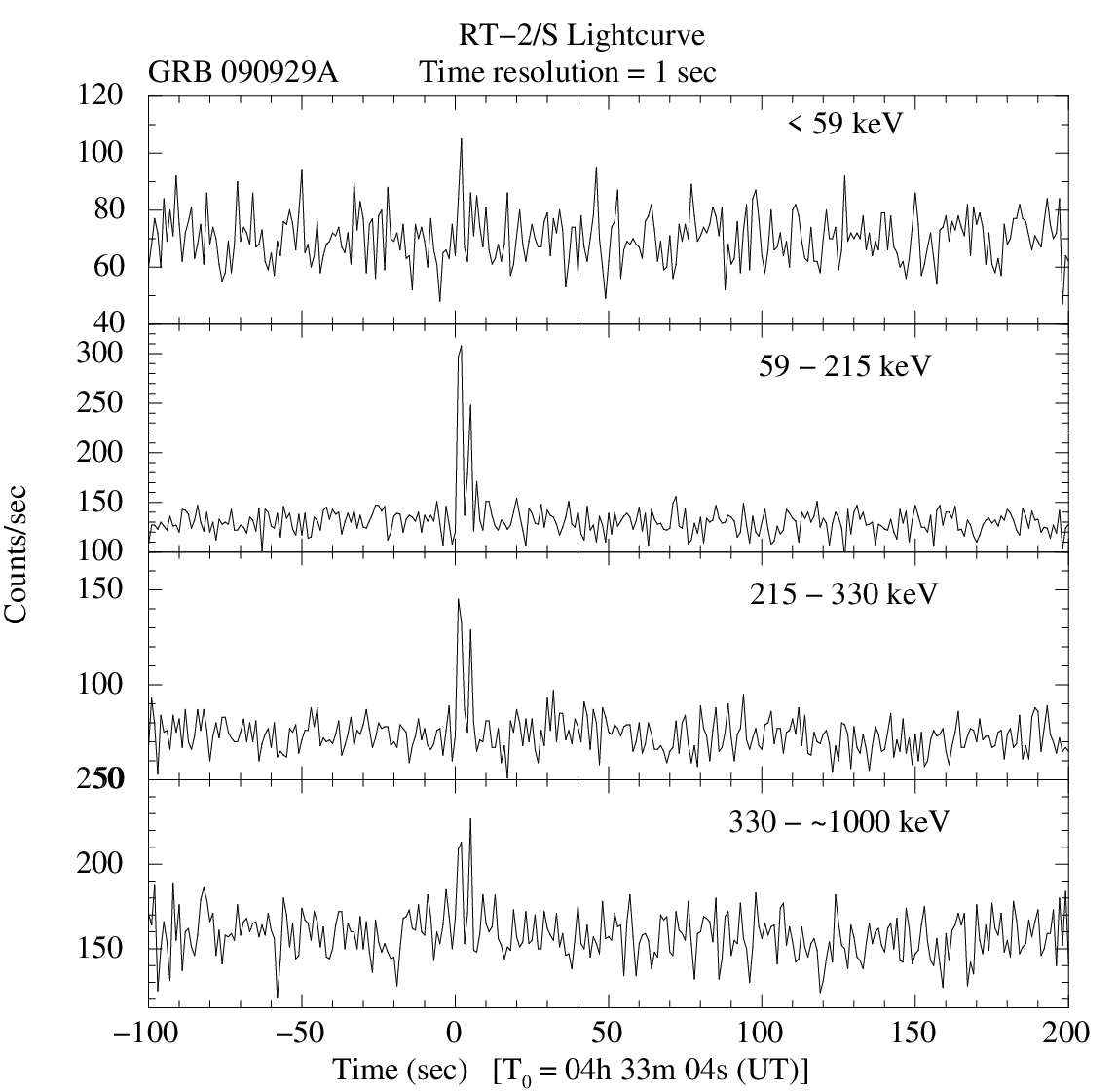}\hspace{0.1 cm}
\includegraphics[height=2.6in,width=2.5in,angle=270]{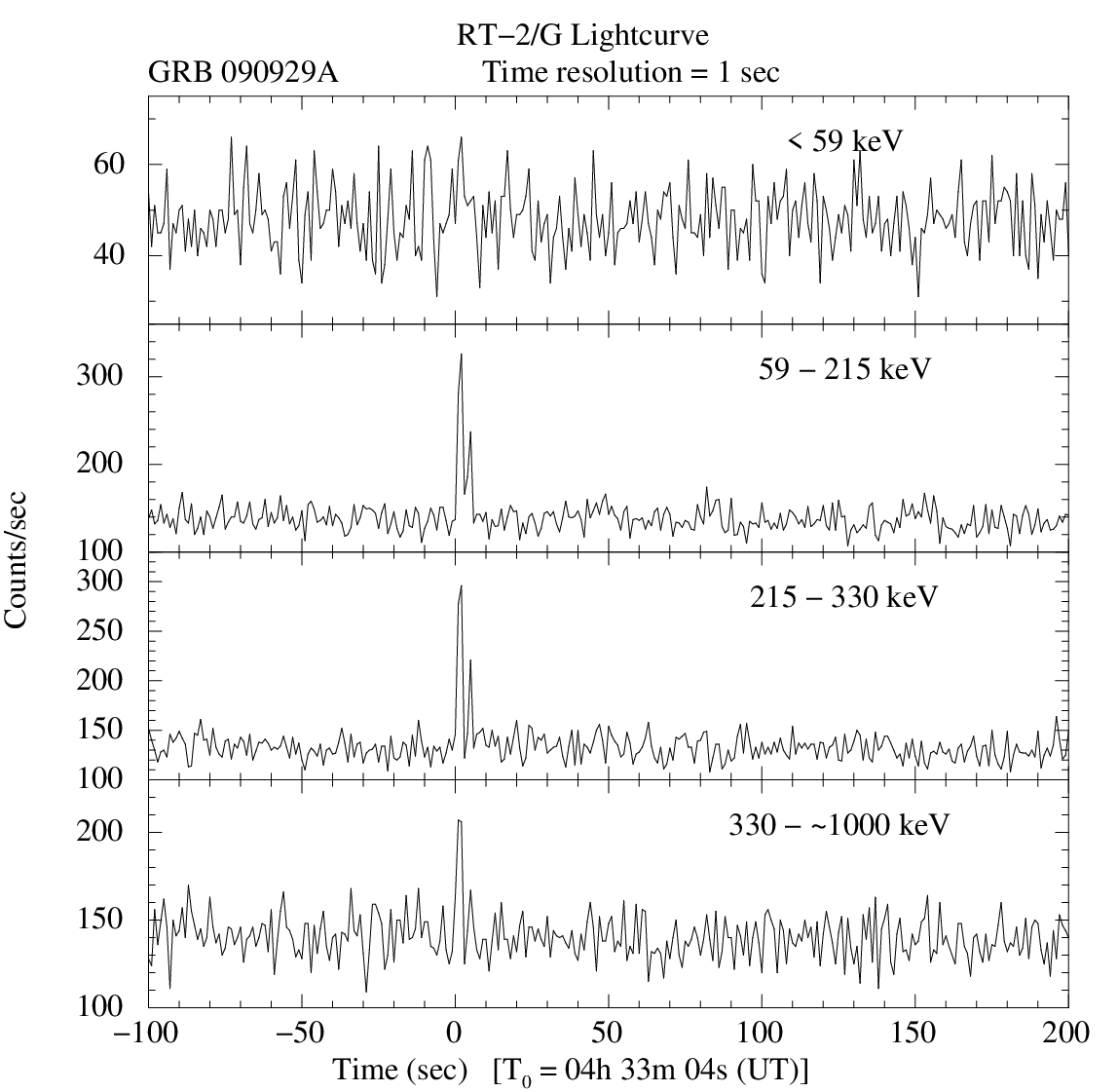}
\caption{(a,b) Light curves in different energy band (marked) of (a) RT-2/S and (b) RT-2/G
detectors. Emission peaks along with the weak precursor trigger pulse at T$_0$ = 04h 33m 04s UT
are detected in a wide energy band (Chakrabarti et al., 2009c).}
\label{kn : fig6}
\end{figure}

Both the phoswich detectors (RT-2/S \& RT-2/G) of the RT-2 Experiment, have also detected
this GRB (Chakrabarti et al., 2009c). During the time of GRB, the satellite was in the LIGHT mode
(pointing towards the Sun) with GOOD time (i.e., away from the polar charge particle and SAA
regions) observation of 1537 sec starting at 04h 21m 20sec (UT) and ending at 04h 46m 57sec (UT).
This burst is also detected by KONUS-RF (Golenetskii et al., 2009c).

The 1 sec binned burst light curves show double peak profile of a total duration
of just around 5 sec. The strongest peak count rate is $\sim 310$ counts/sec in the energy
band of 59-215 keV. In Figure 6., the light curves are plotted at various energy bands for
both the detectors.

%Both RT-2/S and RT-2/G detectors have registered the prompt emission from the
%GRB090929A in the energy band of 60 - 1000, with strongest emission in
%60 - 215 keV energy band.

\section {\bf\large Discussions and Concluding Remarks}

In this paper, we have discussed briefly about the instruments of RT-2 Experiment onboard 
Russian CORONAS-PHOTON mission and discuss the observational results obtained from the 
four GRBs, observed by RT-2. 

Immediately after the commissioning phase of RT-2 payloads, 
GRB~090618 was co-observed by RT-2 along with other X-ray and gamma-ray satellites. A complex, 
multiple peaks were observed in the light
curve of GRB 090618. Spectro-timing analysis showed a systematic softening of the spectrum for 
the successive pulses which is associated with the variations in the timing parameters. 
The measured isotropic energy ($E_{iso}$) and intrinsic peak energy ($E_{p,i}$) of this GRB are 
around $2.2 \times 10^{53}$ ergs and $252$ keV. The results suggest that the GRB 090618 closely
follows the `Amati' relation. It has been recently suggested that GRB090618 has a multiple components,
possibly even the signatures of the black holes (Izzo et al.  2012). %(Ruffini, Chakrabarti, Izzo, 2011; Izzo et al.  2012). 

Timing analysis of other GRBs show that the burst width decreases with the increase of the
energy ranges and also show the nature of the energy dependence of the gamma-ray 
variations. The GRBs that are detected by RT-2, belong to the category of long-duration bursts. 

%For the successive peaks in the GRB, the peak energy ($E_p$) 
%shifts to lower values, the width of the pulse varies sharply with energy and the delay 
%(which is lower for the latter pulses) as a function of energy shows a flatter dependence 
%on energy. 

%%%%%%%%%%%%%%%%%%%%%%%%%%%%%%%%%
\noindent {\bf Acknowledgments}
The RT-2 project was made possible in part from a grant from Indian Space Research 
Organization (ISRO). We also acknowledge other scientific and technical team members of the 
RT-2 project.

%%%%%%%%%%%%%%%%%
%%%%%%%%%%%%%%%%%
\end{document}